\documentclass[journal, 11pt, comsoc, onecolumn]{IEEEtran}

\usepackage{cite}

\usepackage[cmex10]{amsmath}
\usepackage{amssymb}
\usepackage{float}
\usepackage{empheq}
\usepackage{comment}
\usepackage{setspace}
\usepackage[super]{nth}
\usepackage[ruled,vlined]{algorithm2e}

\ifCLASSINFOpdf
   \usepackage[pdftex]{graphicx}
\else
   \usepackage[dvips]{graphicx}
\fi

\usepackage{url}
\usepackage{hyperref}

\usepackage{amsmath}
\allowdisplaybreaks

\begin{document}

\author{Marcelo Ruas,~\IEEEmembership{}%
Alexandre~Street,~\IEEEmembership{Senior Member,~IEEE,} %
Cristiano Fernandes~\IEEEmembership{} %
%Davi~M.~Vallad\~ao,~\IEEEmembership{}%
%Thuener~Silva~\IEEEmembership{}%
%and~Oscar~Dowson~\IEEEmembership{}%
\thanks{M.~Ruas, A. Street, and C. Fernandes are with the Department of Electrical Engineering, PUC-Rio, Rio de Janeiro Brazil.}
}

\title{A Multi-Quantile Regression Time Series Model with Interquantile Lipschitz Regularization for Wind Power Probabilistic Forecasting}

\markboth{}%
{Shell \MakeLowercase{\textit{et al.}}: Bare Demo of IEEEtran.cls for IEEE Journals}

\maketitle

\begin{abstract}
Modern decision-making processes require uncertainty-aware models, especially those relying on non-symmetric costs and risk-averse profiles. The objective of this work is to propose a dynamic model for the conditional non-parametric distribution function (CDF) to generate probabilistic forecasts for a renewable generation time series. To do that, we propose an adaptive non-parametric time-series model driven by a regularized multiple-quantile-regression (MQR) framework. In our approach, all regression models are jointly estimated through a single linear optimization problem that finds the global-optimal parameters in polynomial time. An innovative feature of our work is the consideration of a Lipschitz regularization of the first derivative of coefficients in the quantile space, which imposes coefficient smoothness. The proposed regularization induces a coupling effect among quantiles creating a single non-parametric CDF model with improved out-of-sample performance. A case study with realistic wind-power generation data from the Brazilian system shows: 1) the regularization model is capable to improve the performance of MQR probabilistic forecasts, and 2) our MQR model outperforms five relevant benchmarks: two based on the MQR framework, and three based on parametric models, namely, SARIMA, and GAS with Beta and Weibull CDF.

\end{abstract}


\IEEEpeerreviewmaketitle

%\end{frontmatter}

%\linenumbers


\section{Introduction}

Renewable generation (RG) forecasting is a growing topic among power-systems community due to the number of applications that benefit from it. The intermittent nature of renewable energy sources and the complexity of power-system applications require specific and challenging developments, as for example, the replacement of usual point forecasting methods by more sophisticated probabilistic forecasting approaches. Such a probabilistic forecasts are in general used to improve decisions with risk-analysis-based information relying on the description of extreme quantiles \cite{pinson_probabilistic_2009}. Additionally, according to \cite{Orwig2015}, ``\emph{the uncertainty around
wind and solar power forecasts is still viewed by the power industry as being quite high, and many barriers to forecast adoption
by power system operators still remain.}"

Hence, new statistical models capable of addressing such technicalities have evolved into an emerging field in the power systems literature. See, for example, \cite{bessa2012time,hoeltgebaum2018generating, gallego2016line,moller_time-adaptive_2008,nielsen2006,bremnes_probabilistic_2004,wan_direct_2017,hu2020novel,Gilbert2020}. The main objective in such literature is to propose models capable of generating scenarios from a joint probability distribution of RG time series. This is particularly important in applications in power system based on stochastic optimization models \cite{BirgeBook2011}. For instance, energy trading \cite{Fanzeres2015}, unit commitment \cite{papa2013,papa2015}, grid expansion planning and investment decisions \cite{Aderson2017,munoz2015,moreno2019} are relevant and timely examples where scenarios play a crucial role. 
\subsection{Literature review on wind-power forecast}
In \cite{zhang_review_2014}, the commonly used methodologies regarding wind power probabilistic forecasting models are reviewed, and classified into parametric and nonparametric classes. The main characteristics of \emph{parametric models} are (i) assuming a distribution shape and (ii) low computational costs. The ARIMA-GARCH model, for example, fits the RG series by assuming a Gaussian distribution \emph{a priori}. On the other hand, \emph{nonparametric models} have the following characteristics: (i) do not require a specified distribution for data description, (ii) require more data to fit a model and (iii) have higher computational costs. Popular methods are quantile regression (QR) \cite{gallego2016line,moller_time-adaptive_2008}, kernel density estimation \cite{jeon2012using, bessa2012time},  artificial intelligence \cite{catalao2011short,kariniotakis1996wind}, or a combination of them \cite{he2018probability, wan_direct_2017}.

Although many of the familiar time series models rely on the assumption of Gaussian errors, RG time series, such as wind and solar, are reported as non-Gaussian (see \cite{bessa2012time,jeon2012using} and \cite{Henrique2018}). For instance, in \cite{Henrique2018}, a recent publication proposing a Generalized Auto Regressive Score (GAS) parametric model was proposed and tested to address a wind power time series in monthly basis. In such work, the non-Gaussian model has shown better results in comparison to the traditional based Gaussian-based models such as SARIMA. Still, GAS models (see \cite{creal2013generalized}) rely on the \emph{a priori} choice of the parametric distribution and the estimation process is based on non-convex optimization problems for which global optimality can not be ensured. To circumvent this problem, the usage of nonparametric methods - which do not rely on a distribution assumption - appears as a promising alternative. 

The importance of characterizing the whole distribution becomes even more relevant in applications with asymmetric costs such as those found in power systems applications \cite{Ordoudis2016}. For instance, load-shed costs are in general much higher than the cost of spilling renewable energy; the price and quantity risk due to the need of purchasing in high spot prices when RG falls short of meeting contract amounts is, in general, higher than the risk of clearing RG surplus in low spot prices \cite{Fanzeres2015}; just to mention a few. Hence, having a good estimate of the conditional distribution function (CDF) is essential for meeting accurate estimates of the risk involved in operational and planning decisions. 

The seminal work \cite{koenker1978regression} defines Quantile Regression (QR) as the solution of an optimization problem where the sum of the ``check" function $\rho$ (defined formally in the next session), a piecewise linear convex function, is minimized. Conditional quantiles are the result of the above optimization problem, and this approach is employed in many works \cite{chao_quantile_2012,li_quantile_2007,bosch_convergent_nodate,gallego2016line,moller_time-adaptive_2008,nielsen2006,bremnes_probabilistic_2004,Maciejowska2016a,Maciejowska2016b,wan_direct_2017,Weron2018,Gilbert2020}. Many fields benefit from such applications, ranging from risk measures in asset managements (Value at Risk) to a central measure robust to outliers, but in what follows we will focus on its usage in wind power time series.

% However, by simply joining an array of quantiles
%QR is a tool for constructing a methodology for non-gaussian time series, because of its facility to implement on commercial solvers and to extend the original model.

% However, when estimating a distribution function, as each quantile is estimated independently, the monotonicity of the distribution function may be violated.
% This issue - also known as crossing quantiles - can be adressed by constraining the sequence of quantiles to be in an increasing order. Other possibility is making a transformation afterwards, as shown in \cite{chernozhukov_quantile_2010}.

%
%, as defined in \cite{koenker_quantile_2006}.

%%%%% 4. Falar de regressao quantilica em geral. Onde é utilizada e etc.

% In \cite{koenker_quantile_2006}, the application of QR is extended to time series, when the covariates are lagged values of $y_t$.  
%In our work, beyond autoregressive terms, it is also considered other exogenous variables as covariates. 

%%%%% 5. aplicações de QR em wind power, colocando os papers mais proximos.
% colocar tb regressao quantilica com regularizacao
In \cite{gallego2016line,moller_time-adaptive_2008,nielsen2006,bremnes_probabilistic_2004,wan_direct_2017}, QR is employed to model the conditional distribution of wind power time series. An updating quantile regression model is presented in \cite{moller_time-adaptive_2008}. The authors presented a modified version of the simplex algorithm to incorporate new observations without restarting the optimization procedure. In \cite{nielsen2006}, the authors build a quantile model from an already existent independent wind power forecasts. 

The authors in \cite{gallego2016line} individually estimates multiple quantile regressions, where each quantile model is conceived as a linear regression on a basis of functions. The quantile regression uses regularization through a penalty on the norm in a Reproducing Kernel Hilbert Space (RKHS), which is equivalent to a regularization one the explanatory variables coefficients. This work also develops an on-line learning technique, where the model is updated after each new observation arrives. In \cite{wan_direct_2017}, wind power probabilistic forecasts are made by using QR with a special type of neural network with one hidden layer, called an extreme learning machine. In this setup, each quantile is a different linear combination of the hidden layer features. The authors of \cite{Maciejowska2016a} uses the principal components of a large set of individual forecasting models as covariates in a quantile regression. The Team winning in the electricity price forecasting track of GEFCom2014 proposed a hybrid approach based on pre- and post-processing schemes over the quantile regression averaging approach \cite{Maciejowska2016b}. 
Finally, in \cite{Gilbert2020}, a joint estimation of multiple QR based on gradient boosting trees and spline interpolation was used to generate the predictive CDF. This work shows relevant improvements in the forecasts due to the consideration of turbine-level data.
%Finally, in \cite{hu2020novel}, a joint estimation of multiple QR was proposed based on Mercer’s kernels in the vector-valued and estimated using a metaheuristic, namely, the Multi-Objective Salp Swarm Optimization. The proposed algorithm is capable of mitigating the interquantile issue. %Among others, the post-processing scheme treated the crossing quantile issue and smoothed the final distribution after the QR estimation process.

Variable selection is another topic already explored in previous QR works. The work by \cite{belloni_l1-penalized_2009} defines the properties and convergence rates of QR when adding a regularization term to select covariates according to the Least Absolute Shrinkage and Selection Operator (LASSO) \cite{tibshirani1996regression}. The adaptive LASSO (adaLASSO) used in a QR model was investigated in \cite{ciuperca_adaptive_2016}. In this variant, the penalty for each explanatory variable has a different weight, and this modification ensures that the oracle property is being respected. In \cite{zou_regularized_2008,jiang_interquantile_2014}, the adaLASSO is applied in multiple QR.

Until now, the main benefit of using a multi-quantile regression (MQR) model is the guarantee that the quantiles will not overlap, thereby enabling simultaneously estimated quantile that give rise to a coherent forecasting model for the conditional distribution \cite{wan_direct_2017, hu2020novel}. However, it is well-known that non-parsimonious models, in general, overfit in-sample data sets, and don't provide a good generalization for out-of-sample data. One would also expect a similar behavior in conditional quantile model for the purpose of obtaining a probabilistic forecast. Additionally, global optimality and efficiency are key ingredients to ensure robustness and transparency to estimation processes. For instance, system operators, planners, and market regulators are constantly conducting operation, planning, and regulatory studies that have the potential to affect market prices, system characteristics, and market rules. Therefore, the benefits of parsimonious models and the relevance of computational efficient and transparent estimation methods motivate a step forward in terms of developments in coupled MQR models.

\subsection{Objective and contributions}

The objective of this work is to propose a dynamic model for the conditional non-parametric distribution function (CDF) to generate probabilistic forecasts for a renewable generation time series. To do that, we propose an adaptive non-parametric time-series model driven by a MQR framework. In our approach, the quantile space is discretized within a user-defined granularity and an interpolation method is used to describe a continuous CDF. However, instead of estimating each quantile model separately, all QR models are estimated through a single mathematical optimization problem, which considers relevant features such as smoothness and non-crossing quantiles constraints. In the proposed framework, parsimony is imposed to the coefficient estimates across quantiles and covariates. 

Based on the parsimony principle, we expect that a covariate coefficient should not drastically change within small variations of the quantile probabilities. %The first approach aiming to mitigate this issue was \cite{jiang_interquantile_2014}. In this paper, the absolute value of the discrete first derivative of coefficients across quantiles were penalized within a MQR model applied to cross-sectional data. However, the aforementioned penalization is known to produce stepwise-shaped filtered signals \cite{kim2009ell_1}, contradicting the idea of a single continuous interquantile model for the CDF. 
In this context, an innovative feature of our work with respect to previously reported works is the consideration of a Lipschitz regularization term in the estimation objective function for the first derivative of the estimated coefficients across quantiles. Inspired in the $\ell_1$-filter \cite{kim2009ell_1}, this term penalizes the absolute value of the second-discrete derivative of the QR coefficients in the probability space. As a result, a smooth link among the multiple QR models is induced. A second regularization term, based on adaLASSO \cite{zou_adaptive_2006}, is used to select the best covariates among a set of many candidates. Therefore, the proposed regularization imposes coefficient smoothness, avoiding stepwise-shaped issues (see \cite{kim2009ell_1}), while inducing a coupling effect among quantiles. This coefficient smoothness and induced coupling effect creates the idea of a single (parsimonious) non-parametric CDF model. For the tested data, this model has shown an improved generalization capability in a large rolling horizon out-of-sample test.

Interestingly, in general, statistical estimation procedures focus on minimizing point forecasting errors. However, power-system applications heavily rely on multi-step-ahead probabilistic forecasts \cite{papa2015}.
Unfortunately, universal evaluation metrics summarizing all the characteristics of forecast errors of all qualities are not available. In this context, the evaluation metric should reflect the objective of the user (\cite{Weron2018} and \cite{Pinson2020}). Therefore, to determine the best regularization parameters leading to an accurate probabilistic forecast up to $K$ steps ahead, we test two approaches. In the first one we use a score information criterion (SIC) metric based the QR error measure, which balances the tradeoff between in-sample fit and model parsimony. 
%@@@PARTE ANTERIOR PARA O MAE======================
%In the second approach we optimize a ranked probability score metric based on the conditional probability mean absolute error (MAE) for out-of-sample data. 
%@@@PARTE ANTERIOR PARA O MAE======================
In the second approach we selected the parameters based on the performance exhibited for out-of-sample data. To test the adherence of the probabilistic forecasts (based on estimated CDF's) to unseen data, we applied a rolling-horizon out-of-sample evaluation procedure. As evaluation criterion, we used a probability score metric based on the conditional mean absolute error between the empirical and estimated quantile probabilities.
%% MUDEI AQUI
%the so-called pinball metric, which is consist with the quantile regression objective (see for relevant work adopting the same metric \cite{Weron2018}).

Summarizing, the contributions of this work are twofold:
\begin{itemize}

	\item An adaptive non-parametric CDF-based time-series model for RG. The proposed model is conceived based on a MQR model with two regularization terms. The first term uses an $\ell_1$-filter \cite{kim2009ell_1} to consider a Lipschitz regularization on the first derivative of the coefficients with respect to the quantile probability. The proposed Lipschitz regularization term brings parsimony to the estimation process through a smoothed (continuous) link among the coefficients of the different QR models. The second regularization term considers the adaLASSO method to select the best explanatory variables (auto-regressive terms, exogenous variables, or any function basis).

	\item A linear optimization problem to estimate the proposed MQR-LR model ensuring the aforementioned properties for the predictive non-parametric CDF. The model ensures global optimality in polynomial time to the joint estimation of all parameters. The estimation process is flexible enough to allow 1) a general K-step ahead forecast target and 2) the consideration of constraints (linear inequalities) to the model.  %This is done by the consideration of a $\ell_1$-filter, proposed by \cite{kim2009ell_1}, where the absolute value of the second discrete derivative, across the quantile probabilities, of the coefficients is penalized non-parametric CDF-based time-series model 
	
\end{itemize}

\noindent The features previously described in the two contribution items significantly differentiates our model from the state-of-the-art reported works. As a minor contribution, we perform long-term out-of-sample rolling horizon test to show the improvement of our proposed inter-quantile regularization scheme. For a realistic wind-power generation data from the Brazilian system, results show:  1) the regularization model is capable to improve the performance of MQR probabilistic forecasts in out-of-sample assessments, and 2) our MQR model outperforms five relevant benchmarks, namely, MQR models without any regularization scheme, MQR models without the interquantile regularization scheme, SARIMA models, and GAS models with Beta and Weibull CDF.

The remainder of this work is organized as follows. In Section II, we present the quantile regression framework and the proposed model, the MQR-LR model. In Section III we discuss how to estimate the MQR-LR model in order to obtain a continuous CDF. Section IV shows how to estimate, evaluate and simulate the proposed model. In Section V, we present two computational experiments to evaluate our model: i) a controlled experiment where data is generated through an auto-regressive model and ii) a case study based on real data from a Brazilian wind farm. Section VI concludes this study.

\section{Quantile regression based time series model} \label{sec:qr1}

Let $Q_{Y|X}:[0,1] \times \mathbb{R}^d \rightarrow \mathbb{R}$ be the conditional quantile function of a dependent random variable $Y$ for a given value $x$ of a $d$-dimensional explanatory random variable $X$ (also known as vector of covariates). The $\alpha$--quantile function can be defined as follows:
\begin{equation}
Q_{Y|X}(\alpha,x) = F_{Y|X=x}^{-1}(\alpha) = \inf\{y: F_{Y|X=x}(y) \geq \alpha\}.
\label{eq:quantile-function}
\end{equation}
The function $Q$ is the inverse of the conditional distribution function $F$, and represents the smallest value $y$ for which the distribution function is greater than a given probability $\alpha$.

Let $\rho$ be the ``check" function 
\begin{equation}\label{eq:check-function}
\rho_{\alpha}(x) = \begin{cases}
\alpha x & \text{if }x \geq 0 \\
(\alpha - 1)x & \text{if }x<0
\end{cases}.
\end{equation}
The $\alpha$--dynamic quantile function for $Y_t$ conditioned on $X_t$ can be estimated based on a given finite sample $\{y_t,x_t \}_{t \in T}$, where $T$ is the set of time indexes. The estimation process is given by the solution of the following convex optimization problem:
\begin{IEEEeqnarray}{C}
\hat{Q}_{Y_t|X_t}(\alpha,\cdot) \,\, \in \,\,  \underset{q_\alpha\in \mathbb{Q}_\alpha}{\text{arg min}}\, \sum_{t\in T}\rho_{\alpha}(y_{t}-q_\alpha(x_t)).\label{eq:optim-lqr1} 
\end{IEEEeqnarray}

For inference on QR and finite sample properties, see Chapter 3 in \cite{koenker2005quantile}.
The $\alpha$-quantile function $q(\cdot)_\alpha$ belongs to a function space $\mathbb{Q}_\alpha$. We might have different assumptions for space $\mathbb{Q}_\alpha$ depending on the characteristic we want to consider. A few properties, however, must be part of our choice of space, such as being continuous and having a limited first derivative. However, a general linear regression model, 
\begin{equation}
q_\alpha(x_t) = \beta_{0,\alpha} + \sum_{p \in P}\beta_{p,\alpha} x_{p,t},
\end{equation}
is capable of approximating any well-behaved non-linear function. This can be achieved by expanding the dimension $|P|$ of vectors $\beta_{\alpha} := [\beta_{1,\alpha},...,\beta_{|P|,\alpha}]^T$ and $x_{t}:= [x_{1,t},...,x_{|P|,t}]^T$ to consider as many components as needed of a non-linear functional basis (see \cite{gallego2016line} for an example where a trigonometric basis is used).

% The linear model presumes that the $\alpha$-quantile function is a linear function of its regressors:
% $$q_{\alpha}(x_t) = \beta_{0} + \beta^T x_t.$$   
When dealing with high-dimensional vector of covariates, with many candidates to explain a given quantile, one has to properly select the relevant ones. In practice, this means that some coefficients from the vector $\beta_{p}$ might assume a value of zero, for each quantile $\alpha$. 
There are many ways of selecting a subset of variables among the available options.
A classical approach for this problem is the stepwise algorithm \cite{efroymson1960multiple}, \cite{hocking_selection_1967}, \cite{tibshirani1996regression}, which includes new variables iteratively. 

Newly advocated variable selection methods that fits on a linear programming context are the LASSO/adaLASSO techniques, which consist of penalizing the $\ell_1$-norm of the coefficient's size. In addition to shrinking the coefficients towards zero, it has also the capability of effectively pushing the coefficients to zero (an effect that ridge regression cannot achieve \cite{tibshirani1996regression}). 
The usage of LASSO/adaLASSO in the QR context is the topic of study in \cite{li_l1-norm_2008,ciuperca_adaptive_2016,belloni_l1-penalized_2009,zou_regularized_2008,jiang_interquantile_2014}.
We refer the reader to the work from \cite{belloni_l1-penalized_2009}, where it is possible to find specific properties and convergence rates when using the LASSO to perform model selection in a QR framework. 

Regarding the penalization parameter $\lambda$, which dictates the shrinkage magnitude of the linear coefficients, the level of parsimony of the model can be defined by the user through such quantity. This is because higher values of $\lambda$ means less variables selected to be nonzero. 

The single $\alpha$-quantile adaLASSO is estimated by the following optimization problem:
% {\color{red}
% \begin{IEEEeqnarray}{c}
% \underset{\beta_{0},\beta}{\text{min}} \sum_{t \in T}  \rho_\alpha(y_t - q_\alpha(x_t)) +\lambda \sum_{b \in B} \sum_{p \in P} w_{p} | \beta_{pb}|.\label{eq:l1-qar-optim} 
% \end{IEEEeqnarray}}
\begin{IEEEeqnarray}{c}
\underset{\beta_{0},\beta}{\text{min}} \sum_{t \in T}  \rho_\alpha(y_t - q_\alpha(x_t)) + \sum_{p \in P} w_{p} | \beta_{p,\alpha}|.\label{eq:l1-qar-optim} 
\end{IEEEeqnarray}
What makes the adaLASSO different from the LASSO is the inclusion of the term $w_p$. 
If the model (\ref{eq:l1-qar-optim}) is estimated with all $w_{p}=1$, the output of the optimization problem are coefficients of LASSO  $\beta^{L}_{p,\alpha}$. The adaLASSO coefficients $\beta^{AL}_{p,\alpha}$ are obtained when solving the same optimization problem by letting $w_{p,\alpha}=1/|\beta^{L}_{p,\alpha}|$. 
The main advantage of AdaLASSO over the LASSO is the oracle property for variable selection, which is attended by the former but not by the latter. We refer the interested reader to \cite{ciuperca_adaptive_2016}.

% What differs the adaLASSO from the LASSO is the inclusion of the term $w_p$. Suppose the model (\ref{eq:l1-qar-optim}) is estimated with all $w_{pj}=1$, the output of the optimization problem are coefficients of LASSO  $\beta^{L}_{pj}$. The adaLASSO coefficients $\beta^{AL}_{pj}$ are obtained when solving the same optimization problem while letting $w_{pj}=1/|\beta^{L}_{pj}|$. % para voltar o 'j'

\section{The renewable-generation conditional distribution based on MQR-LR} \label{sec:conditional-distribution}

In the previous section, we presented a linear model for estimating a single $\alpha$-quantile using QR with adaLASSO as a regularization strategy to select the best covariates. However, to build a CDF from an array of quantiles, we propose estimating them at once by a single model in order to explore the coupling effect, i.e., parsimony and generalization capability, across different quantiles.  % This will be discussed in the sequel.

Let the finite discretization of the interval $[0,1]$ be composed of a sequence of probabilities $0 < \alpha_1 < \alpha_2 < \dots < \alpha_{|J|} < 1$ and denote $A$ as the set $A = \{ \alpha_j \mid j \in J \}$, where $J$ is an index set for the probabilities $\alpha$. 
The $\alpha$-quantiles are, from this point forward, indexed by $j$, to account for the different models that are simultaneously estimated. A property that must be respected is the pointwise monotonicity of the estimated quantile function $\hat{Q}_{Y_t|X_t}$, such that $\hat{Q}_{Y_t|X_t}(\alpha_1,\cdot) \leq \hat{Q}_{Y_t|X_t}(\alpha_2,\cdot) \dots \leq \hat{Q}_{Y_t|X_t}(\alpha_j,\cdot)$.
The sequence of quantiles defines a continuous quantile function after interpolation, from which a CDF can be obtained after inverting the estimated quantile function.

To produce a proper distribution function via the estimation of several quantile functions, the estimated quantiles must be monotonically increasing for every $x_t$. If a sequence of quantiles do not respect such a property, this issue is known as crossing quantiles. As we estimate all quantiles at once through a single optimization problem, this issue is explicitly addressed through non-crossing quantile constraints.
In addition to monotonicity, parsimony is a modeling virtue as it mitigates well-known side effects of over-fitting. In our case, where multiple quantile regressions are being jointly estimated to form a single non-parametric CDF, parsimony can be understood as coefficients that do not drastically changes through quantiles. If the coefficient of a given covariate does not follow a smooth profile across the quantile-probability space, it is an evidence that the estimated model is over fitting to in-sample data and might not generalize well the true process. %As a consequence, poor out-of-sample results are expected. To account for all these issues, all quantiles must be estimated at once. 

To ensure parsimonious (smooth) transitions on the estimation of coefficients $\beta_{p,\alpha}$ through the quantiles $\alpha\in A$, we use a $\ell_1$-norm to consider a Lipschitz regularization on the first derivative of coefficients across quantiles. Inspired in second derivative filter \cite{kim2009ell_1}, we define the discrete second derivative of $\{\beta_{p,\alpha}\}_{\alpha\in A}$ as follows:
\begin{equation}
D_{\alpha_j}^{2} \beta_{pj} := \frac{\left(\frac{\beta_{p,j+1}-\beta_{pj}}{\alpha_{j+1}-\alpha_{j}}\right)-\left(\frac{\beta_{p,j}-\beta_{p,j-1}}{\alpha_{j}-\alpha_{j-1}}\right)}{\alpha_{j+1}-\alpha_{j-1}}, 
\end{equation}
where for the sake of notation simplicity, hereinafter we assume $\beta_{p,j}:=\beta_{p,\alpha_j}$. 

Therefore, the proposed MQR-LR model is defined by the vector of coefficients $\beta_{j} := [\beta_{1,j},...,\beta_{|P|,j}]^T$ and intercept $\beta_{0j}$ that solve the following minimization problem:
\begin{IEEEeqnarray}{lr} % para duas colunas
  \underset{\beta_{0j},\beta_j}{\text{Minimize}} \sum_{j \in J} \left( \sum_{t\in T}\rho_{\alpha_j}(y_{t}-(\beta_{0j} + \beta_j^T x_t)) \right. \span \nonumber \\  
  \span \left. + \lambda\    \sum_{p \in P} w_{pj}^\delta \mid  \beta_{pj} \mid \right) + \gamma \sum_{p \in P} \sum_{j \in J'} |D^2_{\alpha_j}\beta_{pj}|, \label{eq:adaLASSO_model_mat1}\\
  \text{subject to:} \span \nonumber \\
	\beta_{0j} + \beta_{j}^T x_{t} \leq \beta_{0,j+1} + \beta_{j+1}^T x_{t},& \forall t \in T, \forall j \in J_{(-1)}, \label{eq:adaLASSO_model_mat2} 
\end{IEEEeqnarray}
where weights $w_{pj} = 1/\tilde{\beta}_{pj}$ are defined based on the values of coefficients $\tilde \beta_{pj}$ estimated with the same model disregarding the the AdaLASSO penalty. As in \cite{medeiros20161}, in this work we set $\delta$ equal to one. The sum of the absolute values that compose the second derivative filter, $\sum_{j \in J'}\sum_{p \in P}|D_{\alpha_j}^{2}\beta_{pj}|$, is added on the objective function multiplied by a tuning parameter $\gamma$. Note that if we consider in the last term of \eqref{eq:adaLASSO_model_mat1} the supremum norm instead (which is straightforward in the linear programming framework), this penalization term would reflect the Lipschitz constant of the derivative of coefficients across quantiles. Finally, $J_{(-1)} = \{ 2, \dots, |J| \}$ is the set containing all indexes but the first and $J' = \{ 2, \dots, |J|-1 \}$ is the set containing all indexes but the first and the last. These two auxiliary sets are used to implement the constraint (\ref{eq:adaLASSO_model_mat2}), which ensures non-crossing quantiles. %As a consequence, a monotonic quantile function is obtained by forcing that, for every $x_t$ and $j$, $q_{\alpha_{j}}(x_t) \leq q_{\alpha_{j+1}}(x_t)$. 
With this approach, an array of regularized (parsimonious) non-crossing quantiles gives rise to a new non-parametric CDF-based time series model. A salient virtue of this model is the coefficient smoothness across quantiles. Based on that, a tighter and endogenous coupling effect among quantiles is induced, which generates the idea of a single CDF model.

\section{Estimation and simulation procedures} \label{sec:estimation-evaluation-simulation}

This section presents computational aspects of the estimation of our proposed model, such as the mathematical programming formulation that accounts for the two regularization terms introduced in previous sections. %We also present the algorithm to perform a Monte Carlo simulation with the estimated model and an out-of-sample procedure to determine the best regularization parameters. 
The methodology is implemented in R \cite{rlanguage2008} and Julia \cite{bezanson2012julia} languages, using  packages JuMP \cite{DunningHuchetteLubin2017}, Gurobi and RCall.

\subsection{Estimation of the MQR-LR model} \label{sec:qral-estimation}

We assume that all covariates are normalized. If they are not in the same scale, the shrinkage feature of the adaLASSO will fail, as different variables may have different weights according to their relative size. Thus, let $\tilde x_{t,p}$ be an input observation at time $t$ of covariate variable $p$.
The normalization process is a linear transformation to each covariate $p$, such that all have a mean of $0$ and a variance of $1$. 
We apply the transformation ${x}_{t,p} = (\tilde x_{t,p} - \bar{x}_{p}) / \hat\sigma_{\tilde x_{p}}$, where $\bar{x}_{p}$ and $\hat{\sigma}_{\tilde x_{p}}$ are the covariate $p$'s unconditional mean and standard deviation, respectively. The response variable $Y$ does not need to be transformed. More information about this process is available in \cite{hastie2014glmnet}.

The MQR-LR model, as described in problem (\ref{eq:adaLASSO_model_mat1})-(\ref{eq:adaLASSO_model_mat2}), can be implemented as a linear programming problem as follows:
{ \small
\begin{IEEEeqnarray}{lr}
	\underset{\beta_{0},\beta,\varepsilon_{t j}^{+},\varepsilon_{t j}^{-}}{\text{Minimize}} \sum_{j \in J} \sum_{t \in T}(\alpha_j\varepsilon_{t j}^{+}+(1-\alpha_j)\varepsilon_{t j}^{-}) \span \nonumber  \\
	% \span + \lambda \sum_{p \in P} \sum_{j \in J} w_{pj} (\xi^+_{pj} + \xi^-_{pj}) \nonumber \\ 
	% \span + \gamma \sum_{p \in P} \sum_{j \in J'} (D2_{pj}^+ + D2_{pj}^-),  \label{eq:adaLASSO-1} \\
	\span + \lambda \sum_{p \in P} \sum_{j \in J} w_{pj} (\xi^+_{pj} + \xi^-_{pj}) 
	 + \gamma \sum_{p \in P} \sum_{j \in J'} (D2_{pj}^+ + D2_{pj}^-),  \label{eq:adaLASSO-1} \\
	\mbox{subject to:} \nonumber & \\
	\varepsilon_{t j}^{+}-\varepsilon_{t j}^{-}=y_{t}-\beta_{0 j}-\beta_{j}^T x_{t},& \forall t \in T ,\forall j \in J,\\
	\xi_{pj}^+ - \xi_{pj}^- = \beta_{pj},&\forall p \in P, \forall j \in J\\ 
	D2_{pj}^+ - D2_{pj}^- = \frac{\left(\frac{\beta_{p,j+1}-\beta_{pj}}{\alpha_{j+1}-\alpha_{j}}\right)-\left(\frac{\beta_{p,j}-\beta_{p,j-1}}{\alpha_{j}-\alpha_{j-1}}\right)}{\alpha_{j+1}-\alpha_{j-1}}, \span   \nonumber \\
	\span \forall p\in P, \forall j \in J',  \\
	\beta_{0j} + \beta_{j}^T x_{t} \leq \beta_{0,j+1} + \beta_{j+1}^T x_{t},&\forall t \in T, \forall j \in J_{(-1)}, \label{eq:qral-crossing} \\
	\varepsilon_{t j}^{+},\varepsilon_{t j}^{-}\geq0,&\forall t \in T, \forall j \in J,\\
	\xi_{pj}^+, \xi_{pj}^- \geq 0, & \forall p\in P, \forall j \in J, \\
	D2_{pj}^+, D2_{pj}^- \geq 0, & \forall p\in P, \forall j \in J'. \label{eq:adaLASSO-ult} 
\end{IEEEeqnarray}
}
Variables $\varepsilon^+_t$ and $\varepsilon^-_t$ represent the quantities $|y-q(\cdot)|^+$ and $|y-q(\cdot)|^-$, respectively. The first line on the objective function in (\ref{eq:adaLASSO-1}) represents the sum of the function $\rho$ over all $j$, i.e., $ \rho_{\alpha_j}(y-q(\cdot)) = \alpha_j \varepsilon^+_{tj} + (1-\alpha_j) \varepsilon^-_{tj}$. The second derivative term $D^2_{\alpha_j}\beta_{pj}$ is implemented on the optimization problem by adding a penalty on the objective function to penalize its absolute value, modeled as the sum of auxiliary variables $D2_{pj}^+ + D2_{pj}^-$. The tuning parameter $\gamma$ controls how rough the sequence of estimated parameters $\{\hat{\beta}_{pj}^{\theta}\}_{j \in J}$ can be given $p$. Note that the whole estimation process is carried out for a fixed vector of regularization parameters $\theta = [\gamma, \lambda]^T$. In this setting, the array of estimated quantiles follows its definition applied to the estimated parameters, i.e., $\hat{Q}_{Y_t|X_t}^\theta(\alpha_j,x_t)
=\beta_{0j}^{\theta} + \beta_j^{\theta T} x_t$.

\subsection{Estimating the regularization parameters} \label{sec:SIC}

To select the best vector of regularization parameters $\theta$, we test two evaluation metrics for a grid of $\theta$ and pick the best according to each metric.

The first metric is based on the score-based information criterion (SIC) applied to the multi-quantile pinball loss function. The quantile regression loss function in \eqref{eq:optim-lqr1} is equivalent to the pinball loss. So, the SIC-based parameter selection metric comprises two relevant properties: (i) it is a modern score metric used in probabilistic forecast and (ii) it considers an information-criterion control for the lack of model parsimony against the maximum in-sample fit objective. For instance, in order for a covariate to be included in the model, it must supply a sufficient goodness of fit.

%\footnote{Related metrics have also been used in other multiple quantile model studies \cite{zou_regularized_2008, jiang_interquantile_2014}.} 

The expression for the SIC metric is as follows:

{ \small
\begin{equation} 
    SIC_\theta = \sum_{j \in J} \left( \log \left(\sum_{t \in T}\rho_{\alpha_j}(y_t - \beta_{0j}^{\theta} - \beta_j^{\theta T} x_t) \right) +  \frac{\log(|T|)|\epsilon_\theta|}{2|T|}  \right),\label{eq:SIC}
\end{equation}
}

\noindent where $\epsilon_\theta$ is the elbow set, defined as $\epsilon_\theta = \{(t,j): y_t - q_{\alpha_j}(x_t) = 0 \}$. The authors in \cite{li_l1-norm_2008} show that the quantity $|\epsilon_\theta|$ is the effective degrees of freedom in the quantile regression.

\begin{comment}

    While the first metric is based on the information criterion, the second approach uses a rolling-horizon scheme to quantify the out-of-sample performance for each value of $\theta$ in the grid search. Then, for each $\theta$, the model is estimated until a given period $\tau$ (resulting in parameters $\{\beta_{0j}^\theta, \beta_j^{T\theta}\}_{j\in J}$) by solving \eqref{eq:adaLASSO-1}-\eqref{eq:adaLASSO-ult} with $T=\{\tau-(H-1),...,\tau-1\}$. Then, based on the out-of-sample data $y_{\tau}$ and estimated quantiles $\{\hat{Q}_{Y_\tau|X_\tau}^\theta(\alpha,x_\tau)\}_{j\in J}$, the one-step-ahead out-of-sample absolute error is computed. Note that $x_\tau$ contains lagged values of $y$. By performing this calculations for a rolling window horizon, $\tau \in T_{RWH}$, the out-of-sample pinball loss is finally defined as follows:
    %
    \begin{equation}
    Pinball_{\theta} = \sum_{j \in J}\sum_{\tau \in T_{RWH}} \rho_{\alpha_j}\Big(y_\tau - \hat{Q}_{Y_\tau|X_\tau}^\theta(\alpha,x_\tau)\Big).
    \label{eq:pinball}
    \end{equation}
    %
\end{comment}

%@@@PARTE ANTERIOR PARA O MAE======================
While the first approach is based on the information criterion, the second approach uses a rolling horizon scheme to quantify the out-of-sample performance for a grid of values of $\theta$. To do that, the mean absolute error (MAE) between the conditional-quantile probabilities, $\{\alpha_j\}_{j\in J}$, and the actual observed frequency of occurrence of each estimated conditional quantiles, $\{F_j^\theta\}_{j\in J}$, is used. It is worth emphasizing that the MAE used in this work measures conditioned-probability errors among all estimated quantiles, thereby differing form the typical way MAE is applied in general point forecast evaluations. In this work, it is calculated as follows: 
\begin{equation}
MAE_{\theta}= \frac{1}{|J|}  \sum_{j \in J}  \left| \alpha_j -  F_j^\theta  \right|.
\label{eq:MAE}
\end{equation}
In \eqref{eq:MAE}, $\alpha_j$ is given by the quantile space discretization, $F_j^\theta$ is computed within a rolling window scheme as follows. For each value of $\theta$, the model is estimated until a given period $\tau$ (resulting in parameters $\{\beta_{0j}^\theta, \beta_j^{T\theta}\}_{j\in J}$) by solving \eqref{eq:adaLASSO-1}-\eqref{eq:adaLASSO-ult} with $T=\{\tau-(H-1),...,\tau-1\}$. Then, based on the out-of-sample data $y_{\tau}$, an indicator function $\delta_{\{y \le \beta_{0j}^\theta + \beta_j^{\theta T} x\}}(y_\tau,x_\tau)$ flags 1 if $y_{\tau}$ belongs to the one-step-ahead forecasted conditional quantile intervals $(-\infty, \hat{Q}_{Y_\tau|X_\tau}^\theta(\alpha_j,x_\tau)]$. Note that $x_\tau$ contains lagged values of $y$. By performing this calculations for a rolling window horizon, $\tau \in T_{RWH}$, $F_j^\theta$ is finally defined as follows:

\begin{equation}
F_j^\theta = \frac{1}{|T_{RWH}|}  \sum_{\tau \in T_{RWH}}  \delta_{\{y \le \beta_{0j}^\theta + \beta_j^{\theta T} x\}}(y_\tau,x_\tau).
\label{eq:Fj}
\end{equation}

\noindent Thus, the best vector of parameters $\theta$ is selected based on the lowest level of the probability MAE metric. %, which is a discrete version of a ranked probability score \cite{Weron2018}. 
In case of assessing the performance $K$-steps ahead, quantiles are obtained by simulation. Finally, it is relevant to mention that depending on the application, it might be interesting to calibrate $\theta$ considering different weights on different quantiles. In this work, however, we will treat every quantile as equal concerning the error measure.
%@@@PARTE ANTERIOR PARA O MAE======================

\subsection{Monte Carlo simulation} \label{sec:scenario-generation}

We use a Monte Carlo (MC) simulation approach to produce a sample of $S$ scenarios, each of which containing a path of up to $K$-step-ahead values for the time series $\{{y}_{\tau,s} \}_{\tau=|T|+1}^{|T|+K}$. Given the estimated model (defined by the selected vector of parameters $\theta$), we build a continuous function $\tilde{Q}_{Y_t|X_t}^\theta(\alpha,x_t)$ that approximates $\hat{Q}_{Y_t|X_t}^\theta(\alpha,x_t)$ for all $\alpha \in [0,1]$ using an interpolation process. Thus, $\tilde{Q}_{Y_t|X_t}^\theta(\alpha_j,x_t)=\hat{Q}_{Y_t|X_t}^\theta(\alpha_j,x_t)$ for all $j \in J$. As we choose the values of $\alpha_j$ so that $0 < \alpha_1 < \cdots < \alpha_{|J|} < 1$, there are no quantile estimates for the intervals $[0,\alpha_1]$ and $[\alpha_{|J|},1]$. These gaps are filled by linearly extending the line that connects $\alpha_1$ to $\alpha_2$ on the left hand side and extending the line that connects $\alpha_{|J|-1}$ to $\alpha_{|J|}$ on the right hand side until the support $[0,1]$ is fully mapped. 

Based on the interpolated quantile function, scenarios for $\{{y}_{\tau,s} \}_{\tau=|T|+1}^{|T|+K}$ by means of a sample $\{u_{s,t}\}_{s,t=1}
^{S,K}$ from a Uniform(0,1) distribution. For the first period ahead ($\tau=|T|+1$), for each $s$, $y_{\tau,s} \longleftarrow \tilde{Q}_{Y_\tau|X_\tau}^\theta(u_{s,\tau-|T|,x_{\tau}})$. Then, for the next periods, $\tau=|T|+2,...,|T|+K$, $x_{\tau,s}$ can be appropriately calculated based on past values $\{..., {y}_{\tau-1,s}\}$. Finally, a new interpolation is carried out based on the $S$ previously generated scenarios to find $\tilde{Q}_{Y_\tau|X_\tau}^\theta(\alpha,x_\tau)$ for all $\alpha \in [0,1]$. Based on the newly obtained quantile function, we can make ${y}_{\tau,s} \longleftarrow \tilde{Q}_{Y_\tau|X_\tau}^\theta(u_{s,\tau-|T|},x_{\tau,s})$ for $s=1,...,S$.

%It consists of first estimating the model for a given time $\tau$ and scenario $s$. Then, for each of these we take the input vector $x_{\tau,s}$ and calculate the discrete quantile function $\tilde{Q}_{y_{\tau}|X}$, which is an intermediate step to estimate the continuous quantile function $\hat{Q}_{y_{\tau}|X}$. This same MC procedure is employed to produce scenario simulations for a variety of MQR models. This procedure is detailed in Algorithm \ref{alg:mc-procedure} that can be found in the Appendix.

\section{Case Studies}

% {\color{red}
% \subsubsection{Benchmark Models}

% We use three different models are used to benchmark the performance of our proposed MQR-LR model, namely:   
% \begin{enumerate}
%     \item \textbf{MQR-B}; 
%     \item \textbf{MQR-LR with $\gamma = 0$} - By its similarity with the proposed model MQR-LR($\lambda,\gamma$), it is an important benchmark to evaluate the inclusion on the smoothing on the coefficients.
%     It is also a particular case of \cite{gallego2016line} with a different regularization norm;
%     \item \textbf{SARIMA} - estimated using the \emph{forecast} \cite{hyndman2008forecastpackage} package from R, via the \emph{auto.arima} function, which selects the best model using AIC. %(details in
% \end{enumerate}
% The tuning parameters $\lambda$ and $\gamma$ for the MQR-LR model are selected according to the two metrics presented in Section  \ref{sec:estimation-evaluation-simulation}: SIC and MAE.

% }

In the following two case studies, our model {MQR-LR} is tested against benchmark models. We study the performance of our model through two case studies. In the first, a controlled study is performed to check the capability of the model to capture well-known patters of Gaussian models. In the second, a real wind-power generation time series from the Brazilian power system is used and the five benchmarks are compared. 

\subsection{Benchmark models}

First of all, it is important to mention that it is beyond the scope of this work to propose a final methodology with the ultimate objective of performing the best probabilistic forecast among all other possible alternative methodologies. Based on our literature review, the use of ensembles and composite of nonlinear forecast models with external variables (also considering climate data) constitute the state-of-the-art (we refer to \cite{Orwig2015, Maciejowska2016b, Pinson2020, Gilbert2020}). Our proposed MRQ-LR is suitable for many of the previously reported features and methods (ensemble of nonlinear models, e.g., based on machine learning methods, nonlinear basis of functions, more sophisticated kernel interpolation, external climate data, etc.). Conversely, the objective of this work is to propose an optimization-based estimation framework (ensuring global optimality) for MQR based on the concept of interquantile regularization.  

Therefore, in the next two subsections, different comparisons are performed to test the modeling capacity and performance of our proposed MQR-LR model. To do that, five relevant benchmarks are used. The first two benchmark are conceived to isolate the effect of the proposed interquantile regularization scheme. In this sense, the first benchmark model is exactly our proposed model but disregarding the two regularization terms in \eqref{eq:adaLASSO-1}, i.e., with $\lambda = 0$ and $\gamma = 0$. Hereinafter, this benchmark model is referred to as \emph{MQR-B1}. To test the effect of the proposed interquantile Lipschitz regularization in our proposed MQR model, we devise a benchmark model keeping the adaLASSO regularization but disregard the second-derivative penalty term (last term) in \eqref{eq:adaLASSO-1}, i.e., similar to our model but with $\gamma = 0$. Therefore, despite of the variety of approaches dedicated to perform explanatory variable selection (see \cite{gallego2016line}, \cite{Gilbert2020}, and references therein), this benchmark is used to isolate the effect of the proposed interquantile regularization scheme from the class of MQR models using regularization terms to perform variable selection. Hereinafter, this benchmark model is referred to as \emph{MQR-B2}.

To cover the family of models relying on parametric CDF's, three other relevant benchmark models are selected. Thus, the third benchmark model is based on \emph{SARIMA} models implemented in \cite{hyndman2008forecastpackage}. Finally, to provide a more interesting benchmark based on state-of-the-art non-Gaussian parametric models, two instances of the Generalized Auto-regressive with Score (GAS) model implemented in \cite{Bodin2020} are selected. Therefore, the forth and fifth benchmark are the GAS using the Beta CDF, referred to as \emph{GAS (BETA)}, and the GAS using Weibull CDF, referred to as \emph{GAS (WEIBULL)}.

\subsection{Controlled Studies I - Auto regressive Process} \label{sec:ar-study}

In our first simulation study, the capability of the proposed MQR-LR model to recover a first-order auto-regressive model, AR(1), is tested and against MQR-B1. In this case, as we know the true model is an AR(1), we define $\lambda=0$ for all MQR models and no variable selection approach is used. Hence, MQR-B1 and MQR-B2 are equivalent and the difference between the MQR-B1 and our MQR-LR relies solely on the interquantile regularization. 

The used AR(1) model is the following:
\begin{equation}
y_t = \beta_0 + \beta_1 y_{t-1} + \varepsilon_t, \quad \varepsilon_t \sim N(0, 1), \quad t=1,\dots,400, \label{eq:ar1}
\end{equation}
with $\beta_0 = 0$, $\beta_1 = 0.3$ and $y_0 = 0$. The interquantile regularization parameter $\gamma$ (see equations (\ref{eq:adaLASSO-1})-(\ref{eq:adaLASSO-ult})) is estimated using cross-validation, which is a popular technique for selecting optimal parameters values in cross-sectional data. 
%Since the simulated process has only one lag, the model selection will not be evaluated, hence $\lambda=0$.
After simulating 1000 different time series given by equation (\ref{eq:ar1}), the three models are estimated.
% TODO referência cross-validation

\begin{figure}
	\centering
	\includegraphics[width=0.7\linewidth]{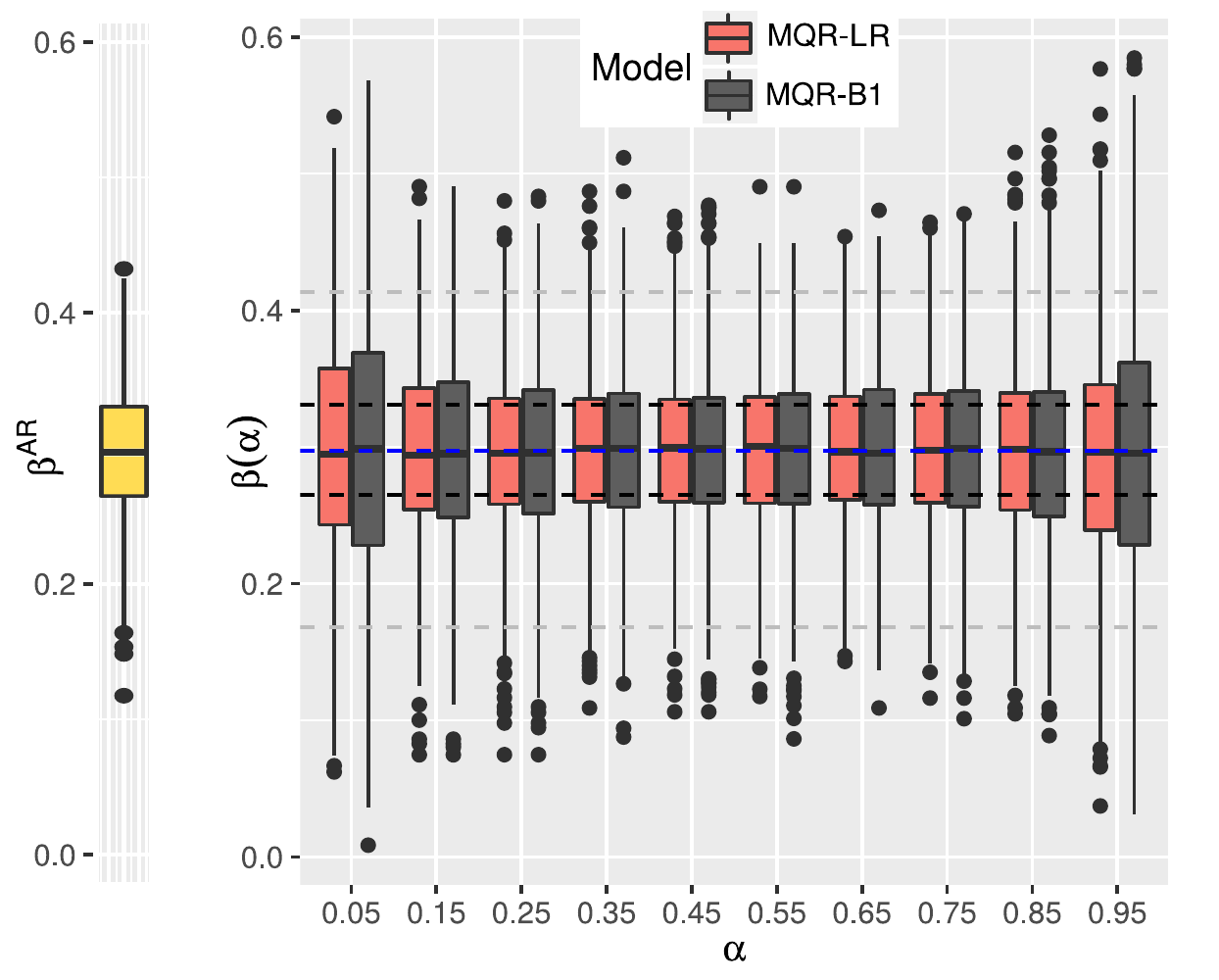}
	\caption{Boxplot showing estimated coefficient after 1000 iterations. The box-plot of the AR(1) coefficient estimation is on the left hand side. Note that for the AR(1) the coefficient is equal for all probabilities $\alpha$. The boxplot of the coefficient estimates for both MQR-B1 and the MQR-LR is on the right hand side.}
	\label{fig:boxplot-ar1}
\end{figure}

Since the main objective of this controlled experiment is to assess the capability of our non-parametric model to recover a given AR(1) CDF, its performance can be evaluated by examining how close the estimated quantiles are from the populational ones. The results for each model are depicted in Figure \ref{fig:boxplot-ar1}, where a box-plot containing the results for the 1000 simulations are shown. %a single boxplot for AR(1) and one for each probability $\alpha$ for QRAL and QRK.
The conclusions from this experiment are: (i) coefficient estimation errors for the central quantiles deviate very little from those estimated by the AR model; (ii) extreme quantiles are usually harder to estimate due to fewer observations avaiable, consequently the estimation error increases on the extremities; (iii) MQR-LR has an advantage over MQR-B1 because it shows smaller variance of estimators.

\subsection{Case study with realistic data}

% \todoi{1) Tira o gráfico do SARIMA (Fig. 7). Vamos dizer apenas assim: na figura 6 mostramos os quantis das distribuicoes simuladas com o modelo QRAL para K=4. Dai comenta rapidamente e pronto. Limitaremos as comparações às tabelas e à fig. 5. 

% 2) Faz caber em 8 páginas. }

In this section, the MQR-LR methodology is tested in generating probabilistic forecasts for a real wind power unit generation. The wind power time series, measured in megawatts, is composed of 2 years (from June-2011 to May-2013) of hourly power generation observations from a wind farm located in the Northeast of Brazil. % Figure \ref{fig:icaraizinho-mensal} depicts a time series plot from which a strong seasonal pattern can be seen. 
% The annual seasonality is seen in Figure \ref{fig:icaraizinho-mensal}, where each individual year is plotted as a single line on the graph. 
% \begin{figure}
% \centering
% \includegraphics[width=0.8\linewidth]{Images/icaraizinho-mensal2.pdf}
% \caption{Icaraizinho annual data. Each series consists of monthly observations for each year.}
% \label{fig:icaraizinho-mensal}
% \end{figure}

As previously mentioned in Section \ref{sec:estimation-evaluation-simulation}, the case study resorts to a rolling horizon scheme. At each step, the model is estimated using a window of size $H=|T|=720$ (approximately one month) and the quantiles of the next $K$ periods are forecasted. The rolling horizon scheme used to evaluate the models and to select the best parameters when using the probability MAE metric is repeated 500 times, i.e., $|T_{RWH}|=500$, each of which comprising a $K$-hour-ahead probabilistic forecast. For the sake of clarity and comparison purposes, all tested models consider the last 48-hour lag information as covariates. %Note that parameters $\lambda$ and $\gamma$ are constant across all data set. Then, the regularization parameters are selected among a thin grid of values according to two metrics: SIC and probability MAE. For the sake of clarity and comparison purposes, all tested models used the same 48-hour lags as covariates.

\subsubsection{Analysis of Results}

We tested our model against the five benchmarks for the one-step-ahead forecast, as is typical in the literature on the theme, and for the four-step-ahead horizon to illustrate the forecast capability of the proposed methodology. The estimation of parametric models remained identical for both one- and four-step-ahead tests. The parameters $\lambda$ and $\gamma$ may differ for each horizon. This guarantees that the model will have the best performance according to the criteria and horizon. %The forecasted quantiles from the MQR-LR model are directly evaluated against the historic data on the one-step-ahead case. On the other hand, when forecasting four steps ahead the CDF is not directly available and need to be empirically obtained. To do that, we use Algorithm \ref{alg:mc-procedure} (Appendix) to simulate scenarios to generate the empirical K-step-ahead CDF for all MQR-based models tested in this study.

Results for our MQR-LR model calibrated with both SIC and MAE are presented in Table \ref{tab:results-icaraizinho}. From the results of this table, it can be seen that the two metrics are effective in improving out-of-sample performance. For instance, note that MQR-B1, where no calibration scheme is needed, out-of-sample evaluation produced a value for MAE equal to 3.5. When calibrating our model via SIC, the out-of-sample evaluation via MAE has decreased to 2.04, which represents a 41.7\% improvement in out-of-sample performance. Notwithstanding, when calibrating our model via MAE, this gain was even higher, as expected. In this case, MAE dropped to 0.98, which represents a 72\% improvement. The same pattern of out-of-sample improvement via MAE is also observed for the 4-hour horizon. 
 
The selection of a particular calibration metric is a modeling choice of the user and should be carefully chosen according to his or her objective. For instance, in this forecasting study, the model calibrated via MAE is outperformed by the model calibrated via SIC when SIC is used as the evaluation metric and vice versa. Notwithstanding, although the metric selection is a modeling choice, we suggest the probability MAE metric as an interesting approach for specifying a robust CDF in terms of performance against unseen data.
 
Additionally, note that the benefit of the two regularization metrics defining the MQR-LR model can be decomposed. Given the MAE drop from MQR-B1 to MQR-B2 (2.09 for the one-hour ahead forecast and 0.12 for the 4h case), we see the benefit of the AdaLASSO regularization, and from MQR-B2 to MQR-LR (0.43 for the one-hour ahead forecast and 0.34 for the 4h case) we can assess the additional benefit of the interquantile regularization.
 
It is worth mentioning that the three parametric benchmarks, SARIMA, GAS (WEIBULL), and GAS (BETA), are outperformed by the proposed MQR-LR calibrated under both metrics (SIC and MAE) in both time horizons.

{
\setlength{\tabcolsep}{2pt}
\begin{table}[h]
	\centering
	\caption{Cumulated statistics across all quantiles}
	\label{tab:results-icaraizinho}	
	\begin{tabular}{llllll}
		\hline
		Model (tuning criteria) & Horizon     & $\lambda$     & $\gamma$     & SIC            & MAE*             \\ \hline
		\textbf{MQR-LR (SIC)}      & \textbf{1h} & \textbf{1}    & \textbf{0}   & \textbf{7.09}  & \textbf{2.04}  \\
		\textbf{MQR-LR (MAE)}      & \textbf{1h} & \textbf{20}   & \textbf{1.0} & \textbf{8.39}  & \textbf{0.98}  \\
		MQR-B2 (SIC)           & 1h  & 0.13   & 0   &  7.75 & 1.65  \\
		MQR-B2 (MAE)           &  1h &  20  & 0   & 8.05  & 1.41  \\
		MQR-B1                      & 1h          & 0             & 0            & 8.34           & 3.50   \\
		SARIMA                   & 1h          & -             & -            & -              & 2.10        \\ 
		GAS (WEIBULL)                  & 1h          & -             & -            & -              &     6.40           \\ 
		GAS (BETA)                  & 1h          & -             & -            & -              &     2.83           \\\hline
		\textbf{MQR-LR (SIC)}      & \textbf{4h} & \textbf{2.5}  & \textbf{0}   & \textbf{13.16} & \textbf{2.03}  \\
		\textbf{MQR-LR (MAE)}      & \textbf{4h} & \textbf{6.75} & \textbf{7.0} & \textbf{13.30}  & \textbf{1.64}  \\
		MQR-B2 (SIC)           & 4h  &  2.5  & 0   & 13.16		& 2.03  \\
		MQR-B2 (MAE)           &  4h & 3.25   &  0  &  13.18  & 1.98    \\
		MQR-B1                      & 4h          & 0             & 0            & 13.73          & 2.10        \\
		SARIMA                   & 4h          & -             & -            & -              &     3.26           \\ 
		GAS (WEIBULL)                  & 4h          & -             & -            & -              &     5.02           \\ 
		GAS (BETA)                  & 4h          & -             & -            & -              &     7.88           \\ \hline

		\end{tabular}
		\\\text{*Probability MAE values are presented in percentage.}
\end{table} 
}

\begin{figure}
	\centering
	\includegraphics[width=0.6\linewidth]{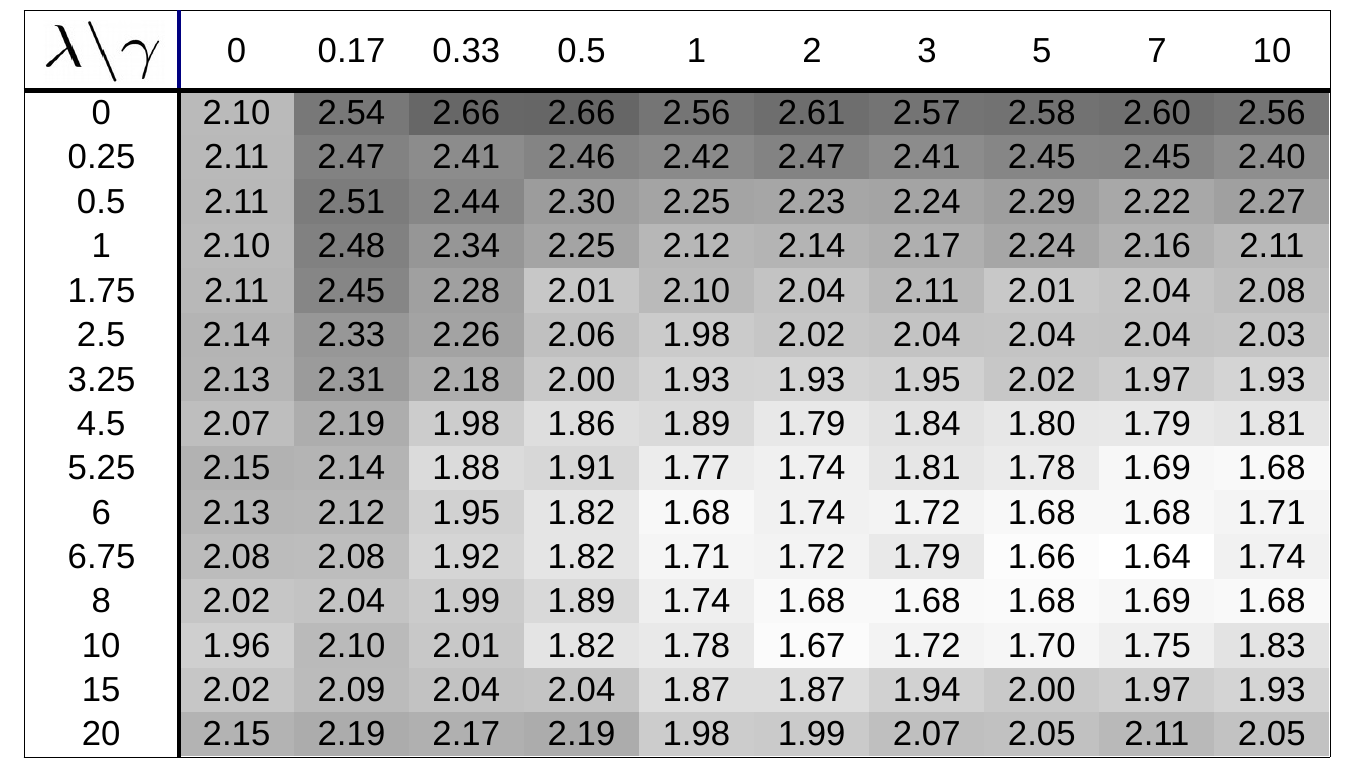}
	\caption{Calculated MAE of forecasting quantiles in a four-hours window. Lower values have a lighter tone, while higher ones are darker. The MAE values are scaled by a factor of 100.}
	\label{fig:heatmap-qral-mae}
\end{figure}

In the sequel, we investigate the forecasting performance of our proposed model for 4 hours ahead. Figure \ref{fig:heatmap-qral-mae} presents a heatmap of the MAE metric for the MQR-LR model considering a combination of regularization parameters. We can see there is a region of optimal regularization levels according to this criteria. The worst performances occur when $\lambda = 0$, such that all covariates are included in the model.

\begin{figure}[h]
	\centering
	\includegraphics[width=0.6\linewidth]{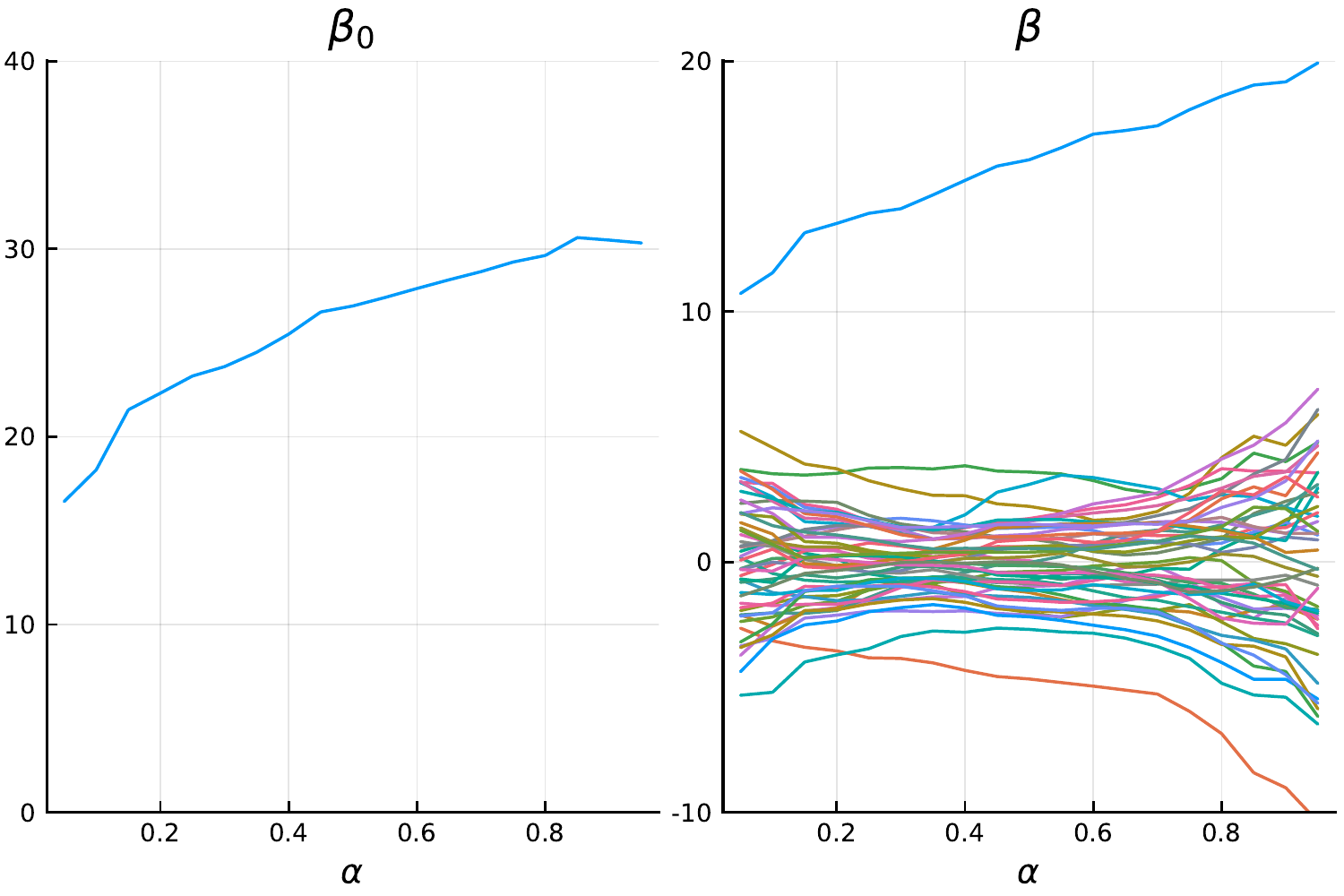}
	\caption{Estimated coefficients for the MQR-B1 model at time $t = 1$.}
	\label{fig:betas-qrk}
\end{figure}

\begin{figure}[h]
	\centering
	\includegraphics[width=0.6\linewidth]{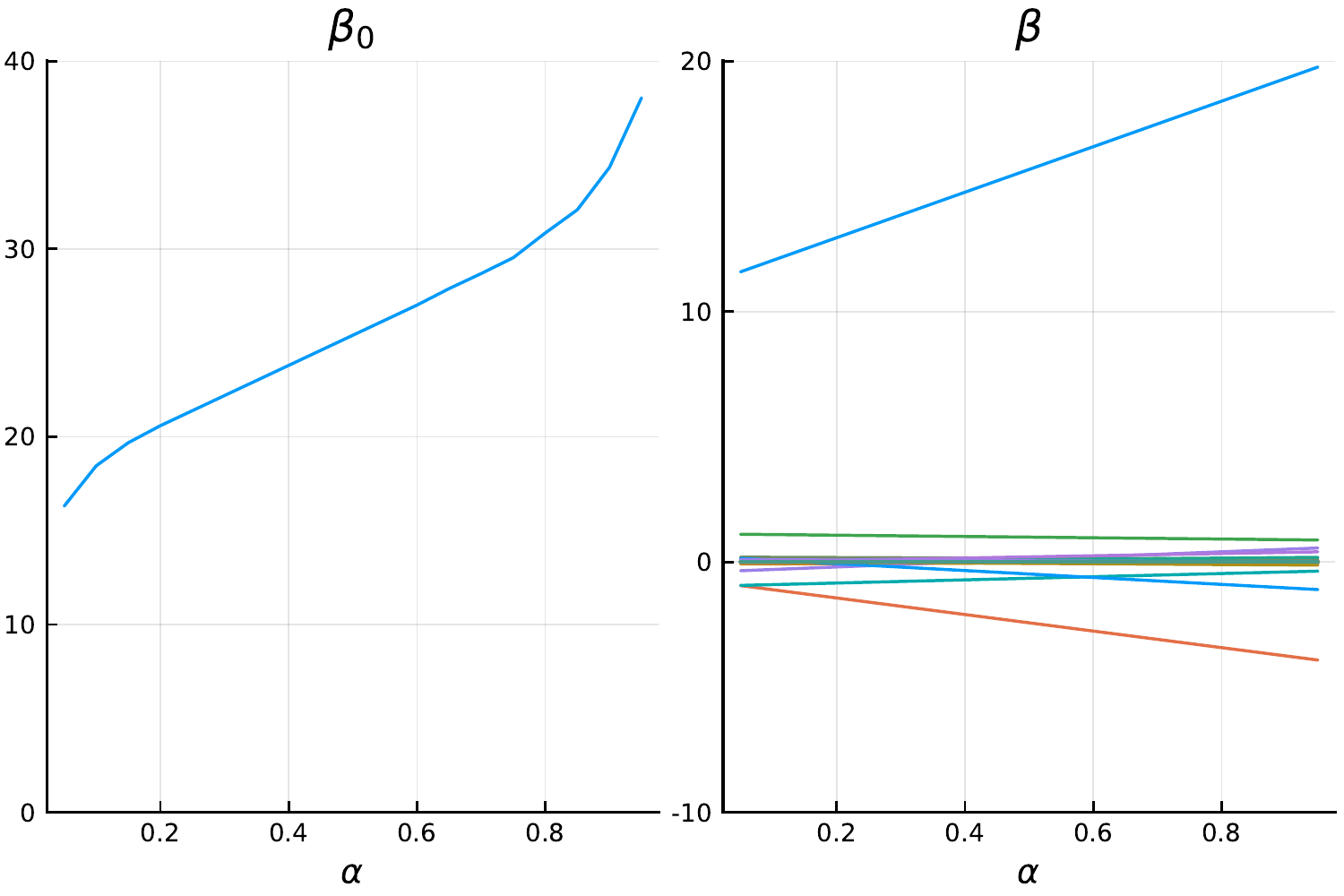}
	\caption{Estimated coefficients for the MQR-LR (MAE) model at time $t =1$.}
	\label{fig:betas-MAE}
\end{figure}

Since coefficients are estimated using a rolling-horizon scheme, they are updated at each step. As the regularization parameters are kept constant, the figures of a given period are representative to understand the coefficients behavior in the experiment as a whole. Figures \ref{fig:betas-qrk} and \ref{fig:betas-MAE} present the estimated coefficients on the first period of the experiment for the MQR-LR (MAE) and the MQR-B1, respectively.
For each model, $\beta_0(\alpha)$ is shown on the left side of the figure, while $\beta(\alpha)$, for each lag, is on the right side. 
The comparison of coefficients across MQR-LR (MAE) and MQR-B1 illustrates the effect of the regularization on the first model.  %MQR-LR is essentially the MQR-B model with regularization on both the variable selection (via LASSO) and interquantile (via the second derivative penalization). 
One advantage of the proposed model is that quantiles are all estimated by a single model, which helps to decrease the variance of the estimators. As a consequence, only a handful of coefficients are selected to be nonzero, and its $\beta$ coefficients follow regularized piecewise linear functions, in contrast to MQR-B1's noisier and higher variability coefficients (as seen in the experiment in section \ref{sec:ar-study}). Note that in Figure \ref{fig:betas-MAE}, although most of the coefficient exhibit a linear pattern, some of them present a piece-wise-linear format. This is due to the section of a regularization norm 1. %If we used a norm-2 instead, our model would become a convex quadratic programming problem for which global optimality would still hold in polynomial time. In this case, smoother curvatures would appear. 

\begin{figure}
	\centering
	\includegraphics[width=0.6\linewidth]{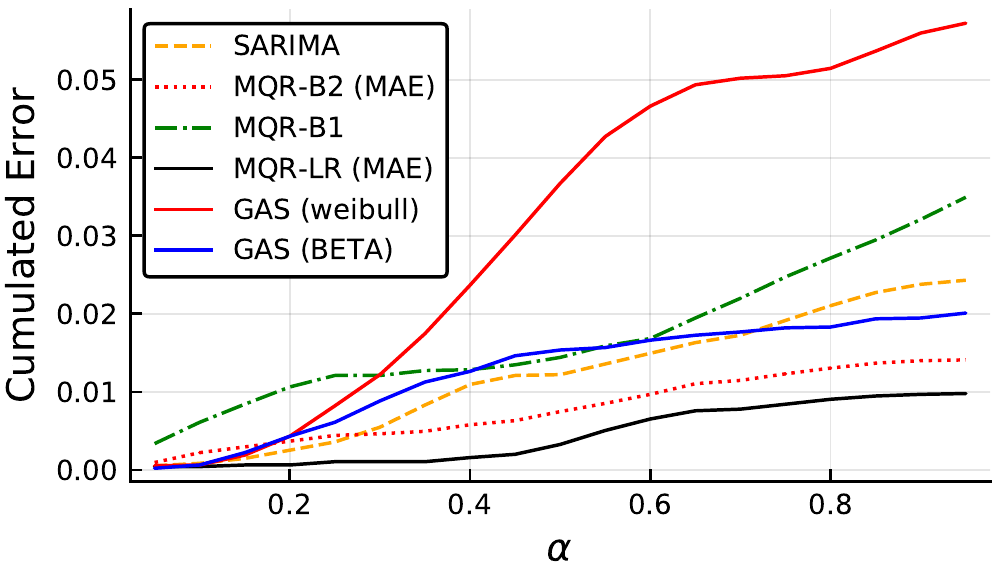}
	\caption{Comparison of empirical probabilities with forecasted one-step-ahead probabilities. The (probability) MAE values are presented in percentage values.}
	\label{fig:prob-prob-plot}
\end{figure}

An advantage of MQR models is the fact that they are able to capture an asymmetric non-Gaussian distribution, which a SARIMA model cannot, as illustrated by Figure \ref{fig:prob-prob-plot}. In this figure we present the cumulative probability error function ($CE$) to compare the distribution fit across quantiles of the MQR-LR model and the benchmarks. %$CE(\alpha_j)= \frac{1}{|J|}  \sum_{i = 1}^j \left| \alpha_i -  F_{i}  \right|$ and it is a partial sum of the $MAE$ function (eq. \ref{eq:MAE}).
A consequence of a better CDF estimation is that simulated scenarios are more accurate in relation to real data, as corroborated by the results presented on Table \ref{tab:results-icaraizinho}. The relatively worse performance of the parametric models (SARIMA and GAS) in contrast to our proposed non-parametric approach revels the challenges of relying on a single class of parametric distribution in this application.    

\begin{figure}
	\centering
	\includegraphics[width=0.6\linewidth]{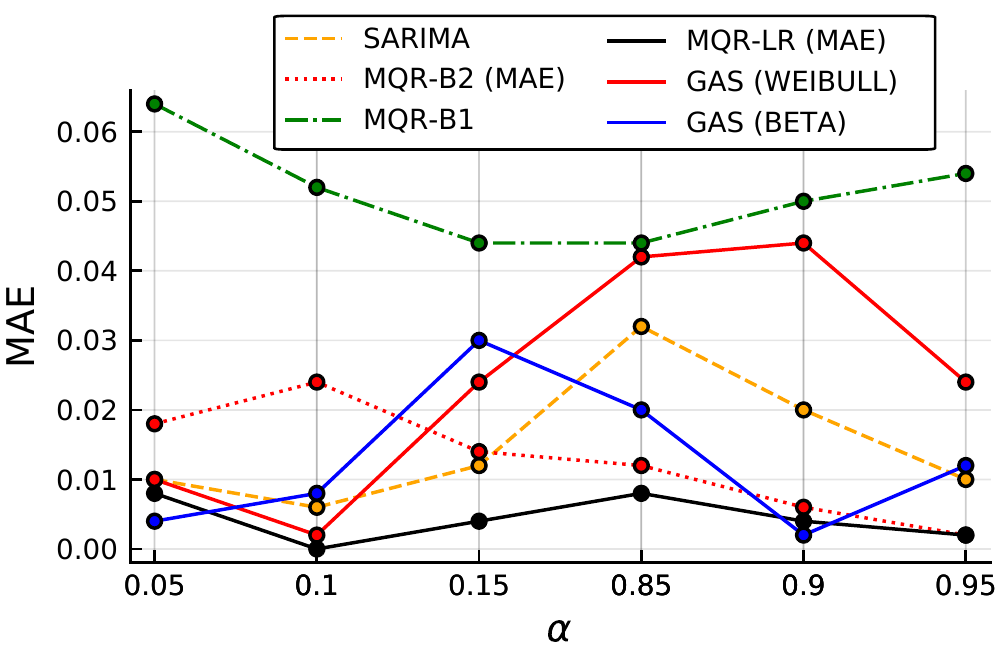}
	\caption{MAE metric restricted to extreme quantiles on the one-step-ahead forecasting. \newline *The MAE values are scaled by a factor of 100.}
	\label{fig:MAE-extreme}
\end{figure}

% \begin{figure}
% 	\centering
% 	\includegraphics[width=1.0\linewidth]{Images/Plot_scenarios-QRK.pdf}
% 	\caption{Comparison of real data with generated scenarios using QRK. The scenarios are generated at the period of each red dot in the plot, with a 4 hours horizon.}
% 	\label{fig:scenarios-qrk}
% \end{figure}
% \begin{figure}
% 	\centering
% 	\includegraphics[width=1.0\linewidth]{Images/Plot_scenarios-SARIMA.pdf}
% 	\caption{Comparison of real data with generated scenarios using SARIMA. The scenarios are generated at the period of each red dot in the plot, with a 4 hours horizon.}
% 	\label{fig:scenarios-sarima}
% \end{figure}
To further illustrate this point and the benefit of the interquantile regularization adopted in our approach, in Fig. \ref{fig:MAE-extreme} we present the probability MAE metric restricted to some extreme quantiles. In this study we account only for the lower and higher quantiles, $\{0.05,\allowbreak 0.10,\allowbreak 0.15,\allowbreak 0.85,\allowbreak 0.90,\allowbreak 0.95\}$. If we compare the MQR-LR with the other MQR benchmarks we see that there is a benefit of connecting extreme quantiles, which rely on few data points, with the more data-reach central quantiles. Based on the proposed idea, extreme conditional quantile can be more accurately estimated.

\begin{figure}
	\centering
	\includegraphics[width=0.6\linewidth]{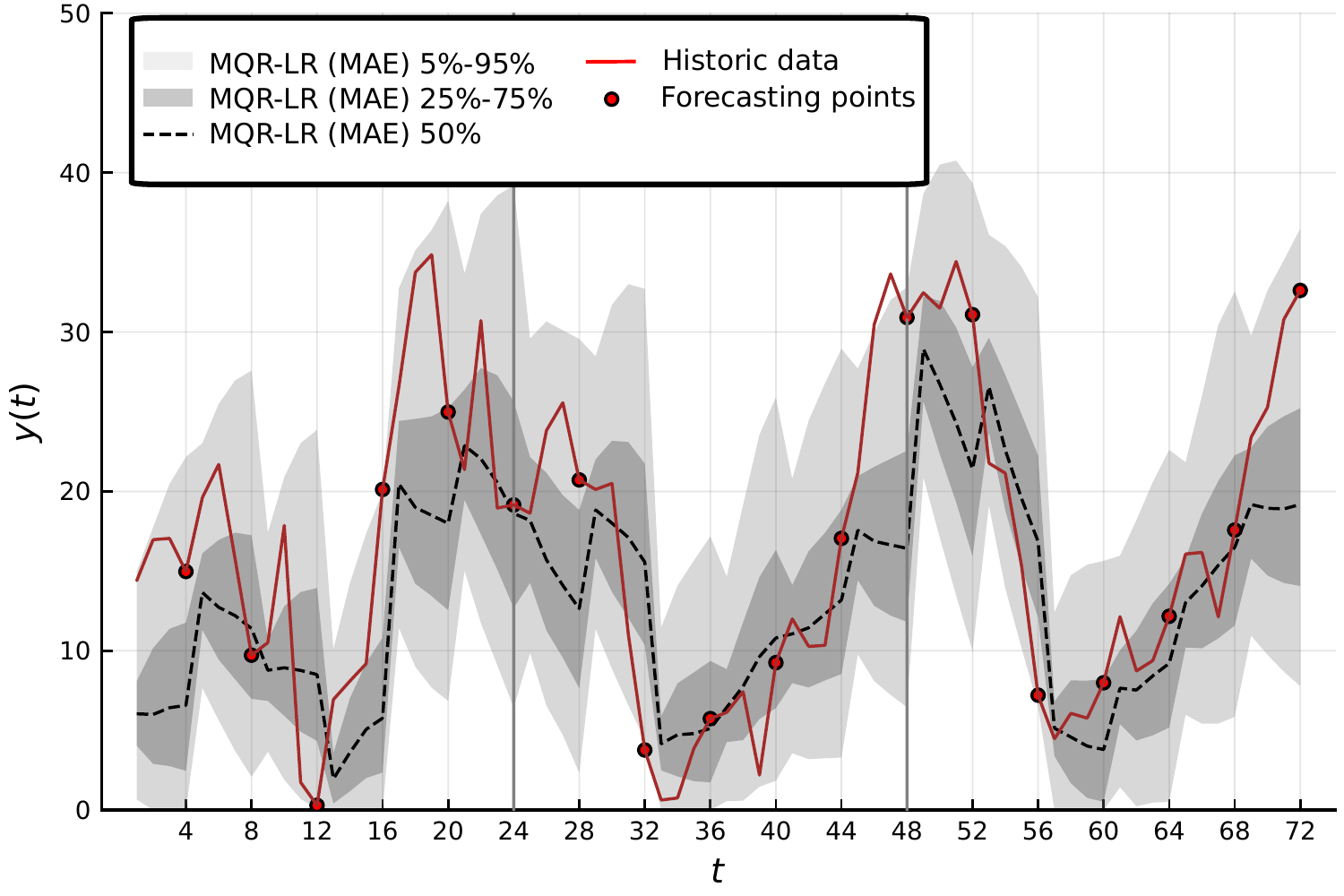}
	\caption{Comparison of real data with generated scenarios using MQR-LR (MAE). The scenarios are generated at the period of each red dot in the plot for the next 4 hours. }
	\label{fig:scenarios-qral}
\end{figure}

Finally, in Figure \ref{fig:scenarios-qral}, we compare the median (50\%), extreme quantiles (5\% and 95\%), and the 1st and 3rd quartiles (25\% and 75\%), obtained from the MQR-LR (MAE) model via simulation,  with the associated historic time series. 

\section{Concluding Remarks}

In this work we propose an adaptive non-parametric conditional distribution function (CDF) time-series model to provide probabilistic forecasts for renewable generation. The model is based on an array of linked quantile regression models. The estimation process simultaneously select all quantile models through a single linear programming problem with two regularization terms, thereby ensuring global optimality to the estimated parameters. The regularization terms account for: 1) explanatory variable selection via adaLASSO, and 2) the link of all quantile models through a Lipschitz regularization applied to the first derivative of coefficients across the quantile probabilities. While the former selects the best covariates, the latter induces a coupling effect capable of improving the performance of probabilistic forecasts. Based on the proposed CDF-based time-series model, we developed an algorithm Monte Carlo simulation algorithm for wind power generation featuring the relevant properties of the empirical distribution and its dynamics. Such a simulation procedure can be used to feed the various applications in power systems relying on simulated scenarios, namely, risk analysis in energy trading, expansion planning, unit commitment, and economic dispatch. 

Our results show that the scenarios generated through the proposed model outperforms five relevant benchmarks such as other Multiple Quantile Regression models, the classical SARIMA models and state-of-the-art GAS models. Furthermore, the linkage effect induced by our proposed Lipschitz regularization scheme is a relevant tool to improve the accuracy of extreme quantiles of the non-parametric CDF. Finally, as interesting future research topics we highlight the consideration of a kernel density estimation process on top of the estimated quantiles and the study of a nonlinear basis of functions.

\ifCLASSOPTIONcaptionsoff
  \newpage
\fi


\vspace{-0.2cm}

\bibliographystyle{IEEEtran}

\bibliography{main.bib}

%\appendix
%\section{Monte Carlo simulation algorithm}

\end{document}